# Searches for HI in the Outer Parts of Four Dwarf Spheroidal Galaxies

L. M. Young

Astronomy Department (MSC 4500), New Mexico State University, P.O. Box 30001, Las Cruces, NM 88003; lyoung@nmsu.edu

## ABSTRACT

Previous searches for atomic gas in our Galaxy's dwarf spheroidal companions have not been complete enough to settle the question of whether or not these galaxies have HI, especially in their outer parts. We present new observations of the dwarf spheroidals Sextans, Leo I, Ursa Minor, and Draco, using the NRAO 140-foot telescope to search much farther in radius than has been done before. The new data go out to at least 2.5 times the core radius in all cases, and well beyond even the tidal radius in two cases. These observations give HI column density limits of $2$–$6 \times 10^{17}$ cm$^{-2}$. Unless HI is quite far from the galaxies' centers, we conclude that these galaxies don't contain significant amounts of atomic gas at the present time. We discuss whether the observations could have missed some atomic gas.

## 1. Introduction

Our Galaxy's dwarf spheroidal companions have long been thought to be old and dead galaxies. They show no evidence for current star formation. Furthermore, searches for HI emission from the dwarf spheroidal galaxies found no evidence of neutral gas (Knapp et al. 1978; Mould et al. 1990; Koribalski, Johnston, & Otrupcek 1994); one exception seems to be Sculptor (Carignan et al. 1998). Optical and UV absorption experiments for Leo I (Bowen et al. 1995, 1997) detected no neutral or low ionization level gas near that galaxy. Thus, it is commonly assumed that the dwarf spheroidal galaxies have no interstellar medium at all.

The traditional picture of dwarf spheroidals has been that they formed their stars in one major star formation event early in the universe, more than 10 Gyr ago. They were then were evacuated of gas by some process and have been evolving passively ever since. The evacuating process could have been stellar winds and supernovae from the first generation of stars, gravitational interactions with our Galaxy, stripping by the halo of our Galaxy, or perhaps other processes.

Recent information on the star formation histories of the spheroidals contradicts this simple picture. Most of the dwarf spheroidals experienced periods of star formation activity at various times from 10 Gyr to 1 Gyr ago (see reviews by Mateo 1998 and Grebel 1999). Fornax and Leo I may even have had star formation as recently as 100 Myr ago (Stetson, Hesser, & Smecker-Hane 1998; Gallart et al. 1999). Since star formation implies the presence of gas, the majority of the spheroidals did contain neutral gas just a few Gyr ago. The history of the spheroidals seems to be more complicated than previously thought. Moreover, there are at least two galaxies which look a great deal like the other spheroidals but which do contain neutral gas: Sculptor (Carignan et al. 1998) and LGS 3 (Young & Lo 1997). Thus, in the light of the detailed photometric evidence for recent star formation, it is important to establish whether or not the spheroidals have a neutral interstellar medium (ISM). Clearly, the presence or absence of gas is an important clue to the history of these galaxies.

The problem is that existing searches for HI in the dwarf spheroidals do not clearly establish whether they have HI or not. Young (1999) summarizes the limitations of previous observations; the major fault is that they covered only a small fraction of the central parts of the galaxies. Thus the HI mass limits, which appear to be very low indeed, do not pertain to the entire galaxy; they pertain only to the small portion of the galaxy which has been observed. In the case of the Draco and Ursa Minor spheroidals (Knapp et al. 1978) a beam with a half-power radius of $5'$ was centered on galaxies whose core radii (semi-major axes) are $9'$ and $16'$. Therefore, less than one third of the area inside the core radii of these galaxies has been observed. Similarly, previous observations of the Sextans dwarf spheroidal (Carignan et al. 1998) centered a beam with a half-power radius of $7.5'$ on a galaxy with a core radius of $17'$. This problem also exists for the Sagittarius dwarf spheroidal, Fornax, and Carina (Koribalski et al. 1994; Knapp et al. 1978; Mould et al. 1990). Substantial amounts of HI could be present in the unobserved parts of the galaxies, or even nearby and outside the optical galaxy. There are many examples of the latter case, including the dwarf irregular M81 dwarf A, where HI is found in a ring outside the optical galaxy (Sargent, Sancisi, & Lo 1983), and Sculptor, where HI is also found outside the optical galaxy (Carignan et al. 1998). In short, it cannot be assumed that HI would have to be in the centers of the dwarf spheroidals, and therefore the existing observations do not constrain their HI contents.

We address these problems by making additional searches for HI emission in the outer parts of our Galaxy's dwarf spheroidal companions. Young (1999) describes interferometric (Very Large Array) observations of Fornax, Leo II, and Draco. The present paper complements that work by using the NRAO 140-foot telescope to search for HI emission in Sextans, Leo I, Ursa Minor, and Draco. These new observations are important because they go out much farther in radius than previous observations. We go out to radii beyond 2.5 times the core radius in all four galaxies. In all galaxies except Sextans, the observations go out to radii beyond or close to the tidal radius. Subsequent sections describe the observations, the detection limits, whether gas could have been missed by the present observations, and the significance of the new data.

## 2. Observations

The observations were made with the NRAO 140-foot telescope during the nights of March 30 to April 11, 1998. The 21cm HFET receiver was employed, and system temperatures were around 19–20 K. The correlator was divided into two banks of 512 channels, each receiving an independent circular polarization. Bandwidths were 2.5 MHz (530 km s$^{-1}$) and 1.25 MHz (264 km s$^{-1}$), which give velocity resolutions of 1.0 and 0.5 km s$^{-1}$. Figures 1 through 4 show the observed locations superposed on the stellar distributions of the galaxies (the isopleth maps of Irwin & Hatzidimitriou 1995). The pointings are spaced by the FWHM of the 140-ft beam, which is 21′.

The observations were made in frequency-switching mode so that any possible HI structures on very large scales would not be removed by position switching. The center positions of all galaxies were observed with 2.5 MHz bandwidth and 1.0 km s$^{-1}$ resolution. Non-central positions for Sextans and Leo I were observed with a 1.25 MHz bandwidth and 0.5 km s$^{-1}$ resolution (and overlapped frequency switching in the case of Leo I), and were later binned to 1.0 km s$^{-1}$ resolution. In all cases, the frequency offset was more than adequate to ensure the inclusion of all possible gas bound to the dwarf spheroidals, which have stellar velocity dispersions on the order of 10 km s$^{-1}$. In order to improve the baselines, observations were made at night and the focus was continually modulated by $\pm 1/8$ wavelength.

The flux calibration scale for this dataset is based on the continuum sources 3C295 and 3C274, which were observed three times per night. They are assumed to have flux densities of 22.22±0.11 Jy and 203.0±6.0 Jy, respectively, at 1408 MHz (Ott et al. 1994). The antenna temperature scale of the telescope was 3.46±0.05 Jy/K when assuming a noise tube temperature of 1.60 K. The effects of atmospheric opacity are ignored in these data; all observations were made at elevations above 27°, so that the maximum opacity correction would be 2.5% (van Zee et al. 1997). Spectral index corrections and the 1% gain effect of focus modulation are also ignored. The pointing was also checked three times per night with observations of 3C274 and 3C295; all pointing offsets were less than 45″ (3.6% of the beam width), so have negligible effect on the data.

After inspection, the individual scans at each position were stacked. Baseline subtraction was done with second order baselines over regions free from Galactic and/or high velocity cloud (HVC) emission. The exception to this statement is Leo I, which was observed with overlapped frequency switching, and third order baselines were used. The resulting spectra are shown in Figures 5 through 8. Table 1 gives the center coordinates and optical velocity adopted for each galaxy in this paper. Table 1 also gives the angular major axis core and tidal radii and the distance of each galaxy, taken from Irwin & Hatzidimitriou (1995).

## 3. Results

Table 2 gives every location observed along with the noise level per 1.0 km s$^{-1}$ channel, integrated intensity, and column density and mass limits. At each position the integrated intensity is summed in a range 60 km s$^{-1}$ wide, centered on the adopted optical velocity from Table 1. The 60 km s$^{-1}$ range is chosen for the following reason: in the absence of other information, we assume that any HI emission would be centered on the optical velocity with a velocity dispersion not very much greater than that of the stars. The spheroidals have stellar velocity dispersions of 13 km s$^{-1}$ and less, and have little rotational support (Irwin & Hatzidimitriou (1995), and references therein). The 60 km s$^{-1}$ range should be wide enough to include any possible emission from the spheroidals, including effects of the width of the HI line and uncertainty in the optical velocity.

The uncertainty in the integrated intensity is a formal estimate which counts two contributions; the dominant one is from the statistical uncertainty of summing noisy channels, and a smaller contribution is from the uncertainty in determining the baseline level. This estimate is described by many authors— for example, Sage (1990). The estimate assumes completely uncorrelated channels (e.g. no baseline wiggles), so it may be an underestimate of the true uncertainty. The HI column density upper limit is three times the uncertainty in the integrated intensity; the advantage of using this estimator is that it provides a conservative limit which is independent of the velocity resolution. The mass upper limit is derived from the column density upper limit and the distance in Table 1.

All of the integrated intensities in Table 2 have significance levels less than $2\sigma$. The column density limits in Table 2 are only a few $\times\ 10^{17}$ cm$^{-2}$, and mass limits are 300–3000 M$_\odot$. Only one spectrum shows a feature which looks as if it might be a real emission line— the southwest position on Ursa Minor. This feature peaks at $-260$ km s$^{-1}$, only 13 km s$^{-1}$ away from the optical velocity of the galaxy. But it is not strong enough to be considered a real detection. The highest possible significance level that this feature can achieve comes from summing five consecutive positive-valued channels, which gives an integrated intensity of 0.054 ± 0.016 K km s$^{-1}$ (3.4$\sigma$). It is unlikely that this feature is anything more than noise. Given the large number of independent pointings in this project, the probability of a noise spike at this level is high; moreover, the feature does not appear in adjacent, more sensitive spectra. In any case, if this feature were real, it would correspond to a column density of 1.3×10$^{17}$ cm$^{-2}$ and a mass of 180 M$_\odot$, and both of these values are well under the limits in Table 2. In short, no HI emission (or absorption) is detected at any position, other than gas that can be attributed to our Galaxy or HVCs.

Direct comparison of the HI mass upper limits in Table 2 to the previously published limits is not straightforward. Such comparisons are complicated by differences in the beam sizes, in the assumed line width (which is not always specified), and in the method used for estimating a detection limit. For Leo I, the mass limit of Knapp et al. (1978) is 5800 M$_\odot$ in a 10$'$ beam, assuming a 15 km s$^{-1}$ line width, after recalculating for the distance assumed in this paper. The current mass limit for the central position of Leo I is 2900 M$_\odot$ in a 21$'$ beam, assuming a 60 km s$^{-1}$ line width. In this case, the present observations give a substantial increase in sensitivity over the previous observations. But it is clear that in all four cases, the major advantage of the present observations is in covering a much larger area than was previously studied.

Earlier HI observations of these four galaxies consisted of one pointing with a half-power radius of 5$'$ to 10$'$. The present data go out to radii of 42$'$ or 21$'$ (Leo I). In all cases the observations cover a region out to a radius of at least 2.5 core radii. For Leo II, Draco, and Ursa Minor, the observations go out to 0.8 to 1.7 times the major axis tidal radius. Sextans, which has a tidal radius of nearly three degrees, is the only case in which the observations do not go out close to the tidal radius; nevertheless, the spatial coverage is still substantially improved.

Table 1 gives, in column 8, the velocity ranges which were observed but which were covered by Galactic or HVC emission. Column 9 of Table 1 gives the velocity ranges that have been searched for HI emission in each galaxy; for Leo I and Sextans the larger velocity range applies to the center position and the smaller range to the non-center positions. We presume that a weak emission line from gas associated with the dwarf spheroidals would not be detectable on top of the strong Galactic and HVC emission. Furthermore, spatial variations in the intensity of the Galactic or HVC emission can be a factor of two and greater on 21$'$ scales, so that position switching or mapping still does not make it possible to detect a weak dwarf spheroidal line on top of strong Galactic or HVC emission. Draco and Ursa Minor are projected onto HVC complex C, which has center velocities around −180 km s$^{-1}$ in this part of the sky (Hulsbosch & Wakker 1988; Wakker & van Woerden 1991). This HVC complex is strongly detected in the present observations and is detectable out to velocities of about $-230$ km s$^{-1}$. Leo I and Sextans are not projected in front of known HVC gas, but Galactic emission covers velocities out to about $+150$ km s$^{-1}$. Fortunately, none of the Galactic or HVC emission extends to the optical velocities of the dwarf spheroidals. Thus the Galactic and HVC emission probably does not hide gas which is associated with the dwarf spheroidals, unless the optical velocities are wrong by 20 to 140 km s$^{-1}$.

## 4. Discussion

No HI emission was detected from the dwarf spheroidals, with column density limits of 2–6×10$^{17}$ cm$^{-2}$ and with mass limits of 300 to 3000 M$_\odot$. To understand the properties of atomic gas which could escape detection, consider the observed quantity: the brightness temperature (after continuum subtraction), $\Delta T_B$. The brightness temperature can be related to the gas's physical parameters of spin temperature $T_{\rm spin}$, optical depth $\tau$, and beam filling factor $\Phi$ in the usual way

$$\Delta T_B = \Phi(T_{\rm spin} - T_{\rm bg})(1 - e^{-\tau}).$$

In the present case the background brightness temperature $T_{\rm bg}$ refers to the microwave background at 2.7 K. Thus, there are three ways that atomic gas could escape detection by the present observations: it could have very low optical depth, hence low column density; small beam filling factor; or spin temperature very close to 2.7 K. These three possibilities are discussed below, and we argue that none of these possibilities is likely to have hidden a substantial mass of atomic gas.

The column density detection limit is derived from integrating the observed brightness temperature over a velocity range, and therefore the column density limits in Table 2 refer to the product of the true gas column density and the unknown beam filling factor $\Phi$. HI emission could be present at the observed positions with column densities greater than $10^{17}$ cm$^{-2}$, provided that the gas is patchy. For example, a circular source of diameter $2'$ would fill only 1% of the beam, and such a source could escape detection in our data even if the column density were as high as $10^{19}$ cm$^{-2}$. But because the gas mass is integrated over area, the gas mass limits in Table 2 are independent of the beam filling factor. Also note that this kind of bright, patchy emission would be easily detectable in interferometric observations. The VLA observations of Draco, Leo II, and Fornax (Young 1999) provide direct evidence that bright, patchy HI emission is not present in those galaxies. These complementary data, along with the fact that the mass limit is independent of beam filling factor, suggest that there isn't (much) HI which has escaped the present single-dish observations by virtue of small beam filling factors.

HI with very low column density (low optical depth) probably doesn't exist in the dwarf spheroidals; it is expected to be ionized. Sensitive observations of a spiral galaxy (van Gorkom 1993) show that the HI disk of the spiral cuts off sharply when the HI column density reaches about $10^{19}$ cm$^{-2}$. A similar effect is seen in high velocity clouds, where Colgan et al. (1990) observe a tendency for the HI in the clouds to cut off sharply at column densities below $5\times10^{18}$ cm$^{-2}$. Corbelli & Salpeter (1993a, 1993b) and others have argued that these HI cutoffs are caused by ionization by the galactic and/or extragalactic UV radiation field. In fact, ionized gas outside of the truncated HI disk has been detected in NGC 253 (Bland-Hawthorn, Freeman, & Quinn 1997; Bland-Hawthorn 1998). In this picture, the dwarf spheroidals are probably similar to high velocity clouds, and hydrogen at column densities below about $5\times10^{18}$ cm$^{-2}$ would be photoionized by our own Local Group galaxies or quasars. Therefore, it is unlikely that the present observations could have missed substantial amounts of either low column density HI with large beam filling factors or high column density HI with small beam filling factors.

The third way in which atomic gas could evade detection is by having a spin temperature close to 2.7 K. The true column density could then be greater than the apparent column density (which is based on the assumption that spin temperatures are not close to 2.7 K). Current data cannot rule out this possibility, but it is not usually considered to be likely. Corbelli & Salpeter (1993a) studied the excitation of HI in an environment of low pressure and low heating rate— conditions which were intended to be representative of the far outer disks of spirals, and which are probably applicable to the dwarf spheroidals as well. The result is that the extragalactic UV background, via Lyman $\alpha$ pumping, is expected to keep the spin temperature of HI well above 2.7 K even when collisional pumping is minimal. In the most extreme case of an unrealistically low UV flux, Corbelli & Salpeter find that the spin temperatures can drop low enough so that the true column density is four times greater than the apparent column density. In more realistic cases, the difference is much less than a factor of four. Again, this mechanism does not seem to be a viable method for hiding large amounts of atomic gas.

Of course, it is always possible that atomic gas could be outside the regions observed. In Sextans, HI emission could be located between a radius of $42' = 2.5$ core radii and the tidal radius. For the other three galaxies, a significant amount of atomic gas could remain undetected only if it was outside the tidal radius. A search for this kind of "extra-tidal" gas would require sensitive surveys, which are not presently planned for the northern hemisphere. In fact, Lin & Murray (1999) have proposed an interesting scenario in which gas might be associated with the spheroidals, outside the tidal radius, and strung out along the spheroidals' orbits. At some later apogalactic point, the gas could be recollected into the dwarf galaxy proper. But in this picture, the co-orbiting gas is very low column density and is ionized, not neutral. As possible evidence against this picture, note that the absorption experiments of Bowen et al. (1995, 1997) give column density limits of $10^{15}$–$10^{16}$ cm$^{-2}$ for neutral or low ionization state gas outside the tidal radius of Leo I. Nevertheless, except perhaps for Sextans, it is unlikely that substantial amounts of *neutral* gas would be associated with the dwarf spheroidals and outside the regions observed.

Finally, it must be conceded that substantial amounts of neutral gas could be hidden in the dwarf spheroidals in molecular form. Molecular hydrogen itself would be difficult to detect except in the UV absorption lines, and low metallicities might make the CO tracer molecule virtually undetectable as well. However, one might propose to detect molecular clouds in the spheroidals via their HI emission; after all, molecular clouds in our own Galaxy are associated with HI envelopes of column density $10^{20}$ cm$^{-2}$ and higher (Blitz & Williams 1999). But molecular clouds in the dwarf spheroidals would probably not have such extensive HI envelopes. Theoretical models (e.g. Draine & Bertoldi 1996) explain that the depth of the atomic envelope around a molecular cloud depends on the ratio of the local UV field strength to the gas density. The Galactic molecular clouds have HI envelopes of $10^{20}$ cm$^{-2}$ and higher because they are bathed in relatively strong UV fields. But if the molecular gas density could be maintained at 100 cm$^{-3}$ or higher (a value which is typical for giant molecular clouds in our Galaxy's disk), and if the UV field (at 1000 Å) is one tenth as strong as the standard "solar neighborhood" or Habing field, molecular clouds in the spheroidals could have atomic envelopes of column density $10^{18}$ cm$^{-2}$ and less. The more exotic, extremely dense ($10^9$ cm$^{-3}$), fractal molecular clouds postulated by Pfenniger, Combes, & Martinet (1994) would have even smaller atomic envelopes. Thus, ignoring questions of the formation of $H_2$, it is easy to see that a substantial mass of molecular gas could be present in the dwarf spheroidals, undetected by existing observations.

## 5. Implications

We have argued that it is unlikely that the dwarf spheroidals contain substantial amounts of atomic gas, unless that gas is quite far from the centers of the galaxies. Some interstellar medium could be hidden in the galaxies in ionized or molecular form. This situation is an important facet of what Mateo (1998) calls "the interstellar medium 'crisis' " in dwarf spheroidal galaxies. Some of the spheroidals show evidence for very recent star formation, and they might be expected to have interstellar gas. Even in the spheroidals without recent star formation, evolved stars should be pumping gas back into the system in detectable amounts.

Of the four galaxies studied in this paper, Ursa Minor and Draco most closely conform to the traditional view of dwarf spheroidals as "Population II" systems. Color-magnitude diagrams for Ursa Minor and Draco indicate that the bulk of the stars are old, $> 10$ Gyr, with very small, if any, intermediate age or young stellar populations. Recent work on the color-magnitude diagrams of these galaxies has been published by Martínez-Delgado & Aparicio (1999) and Grillmair et al. (1998). The Sextans dwarf spheroidal also appears to be predominantly old, though up to 25% of the stars may have intermediate ages of a few Gyr (Mateo et al. 1995). The most striking of the four is Leo I, in which the majority of the stars are intermediate-age stars (1–10 Gyr). Few of the stars are older than 10 Gyr, and some of the stars may be as young as a few hundred Myr (Gallart et al. 1999; Caputo et al. 1999). Star formation implies that Leo I did contain a substantial amount of cold, neutral interstellar gas just a few Gyr ago, and maybe just a few hundred Myr ago. The same is true for Fornax, in which HI has also not been detected (Stetson et al. 1998; Young 1999). The very recent star formation in these galaxies and the apparent lack of any substantial neutral medium are puzzles which we cannot resolve at this time.

Furthermore, in all four spheroidals observed here, mass loss from evolved stars might be expected to exceed the gas detection limits in relatively short times. According to Faber & Gallagher (1976), the mass loss rate in an old stellar population should be about 1.5 $M_\odot$ yr$^{-1}$ per $10^{11}$ $L_\odot$. The V-band luminosities of the present sample of spheroidals are 3–5×$10^5$ $L_\odot$ for Ursa Minor, Draco, and Sextans, and 5×$10^6$ $L_\odot$ for Leo I (Mateo 1998). If all the gas lost from evolved stars is in the form of atomic gas, and all of it settles to the central pointing, this gas would exceed the present detection limits in only $4 \times 10^7$ to $10^8$ yr. If the gas is extended over the entire surveyed area, it would exceed the present detection limits in $2 \times 10^8$ to $2 \times 10^9$ yr. Again, what has happened to this gas is a puzzle. If the gas from evolved stars has high enough density, perhaps it is hidden in molecular form; if it has low enough density, perhaps it is ionized; perhaps this gas has been completely removed from the galaxies.

Have the spheroidals always contained gas, but in forms which have not yet been observed? Do the spheroidals occasionally acquire neutral gas (perhaps from the Magellanic Stream or high velocity clouds), spend part of their lives looking like Sculptor and/or LGS 3, and later lose their gas? One thing is clear: despite the variety of interesting hypotheses about the evolution of the dwarf spheroidals, we do not yet understand their histories, and this problem remains one of the outstanding problems in the evolution of small galaxies and loose groups.

## 6. Summary

We have searched for HI emission in and aound our Galaxy's dwarf spheroidal companions Sextans, Leo I, Ursa Minor, and Draco. New information on the star formation histories of dwarf spheroidals shows that many of them formed stars a few Gyr ago, and some as recently as 100 Myr ago. Furthermore, previously published observations of the dwarf spheroidals covered only a small fraction of the galaxies' areas, so they could not exclude the possibility of substantial amounts of gas elsewhere in the galaxies. The present observations cover much larger areas than previously studied; however, no HI was detected in these galaxies, down to column density limits of $2$–$6\times10^{17}$ cm$^{-2}$. From these observations we conclude that there is no significant HI within the tidal radius of Leo I, Draco, or Ursa Minor, or within 2.5 core radii for Sextans.

Many thanks to J. Gallagher for suggesting the idea of this project and for providing invaluable advice. Thanks to M. Irwin for providing the isopleth maps of the dwarf spheroidals, and to B. Wakker for information on high velocity clouds.

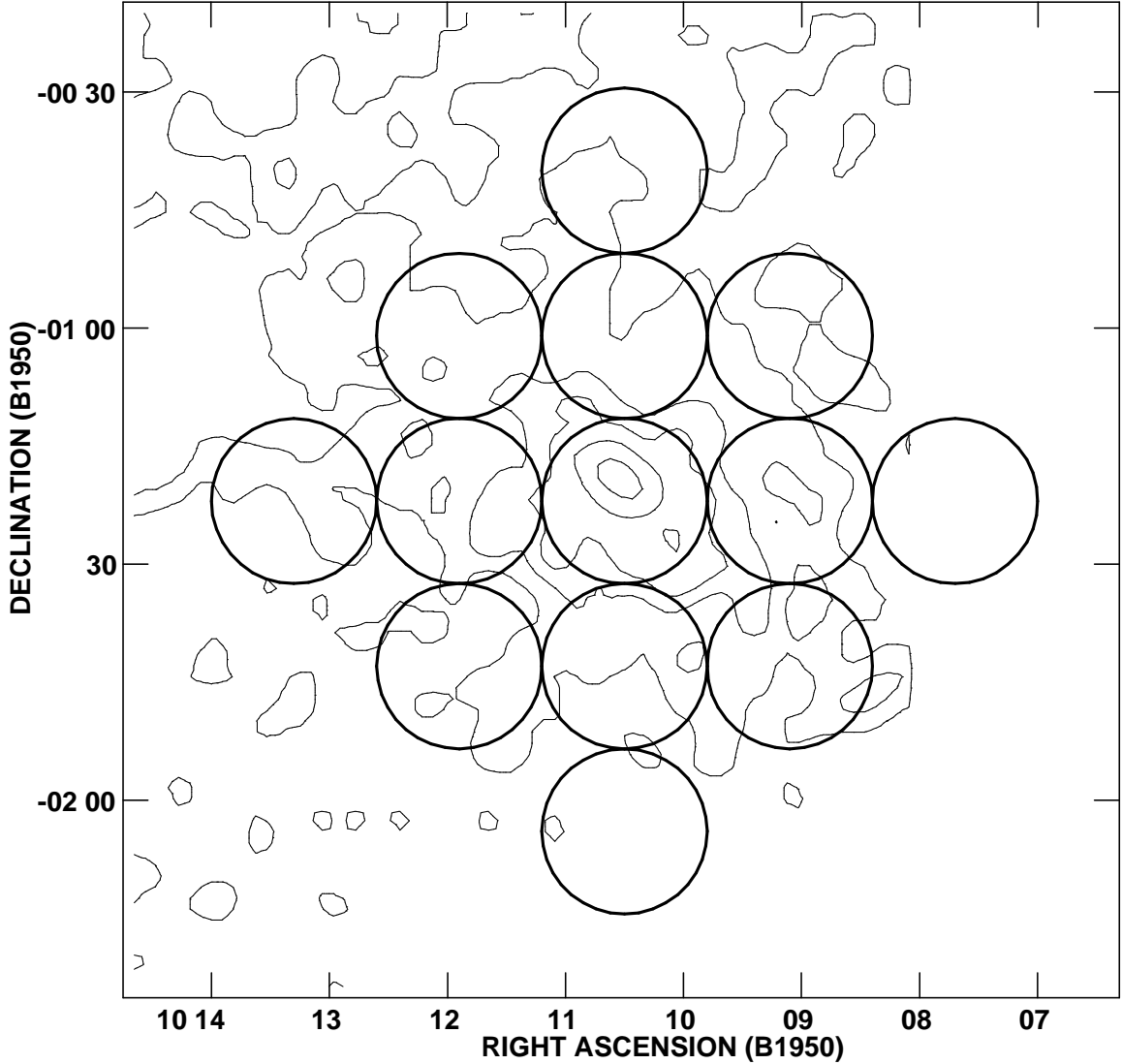

Fig. 1.— Optical extent of the Sextans dwarf spheroidal compared to the locations observed with the 140ft telescope. The light contours show the isopleth image of Sextans from Irwin & Hatziditmitriou (1995). The contour levels are described in that paper. The heavy circles show the locations observed at the 140ft telescope; the circles are the size of the beam full width at half maximum. The most distant observations are 42′ from the galaxy center, which is 2.5 times the core radius or 0.3 times the tidal radius.

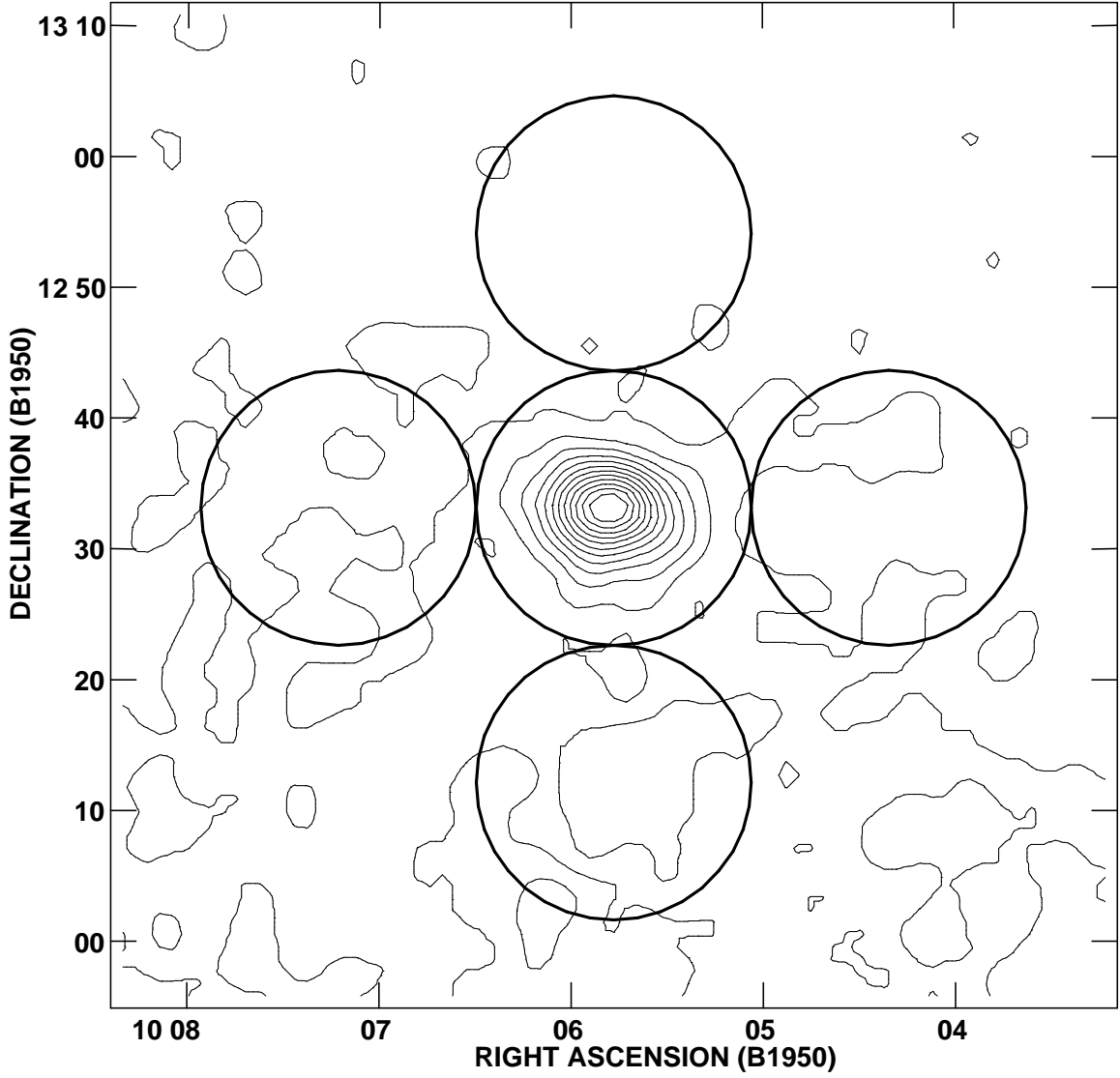

Fig. 2.— Same as Figure 1, for Leo I. The most distant observations are 21′ from the galaxy center, which is 6.4 times the core radius or 1.7 times the tidal radius.

– 15 –

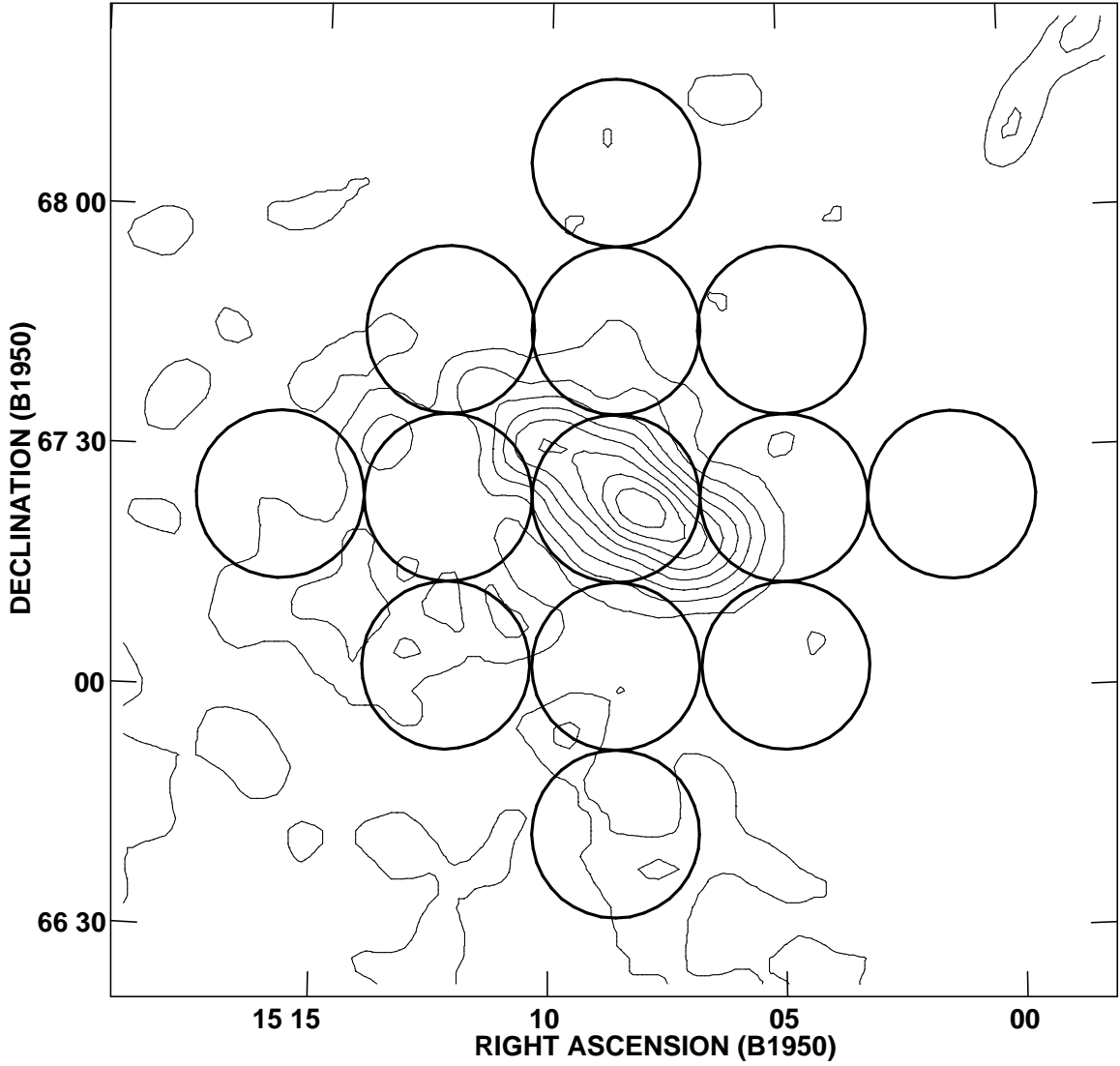

Fig. 3.— Same as Figure 1, for Ursa Minor. The most distant observations are 42′ from the galaxy center, which is 2.7 times the core radius or 0.8 times the tidal radius.

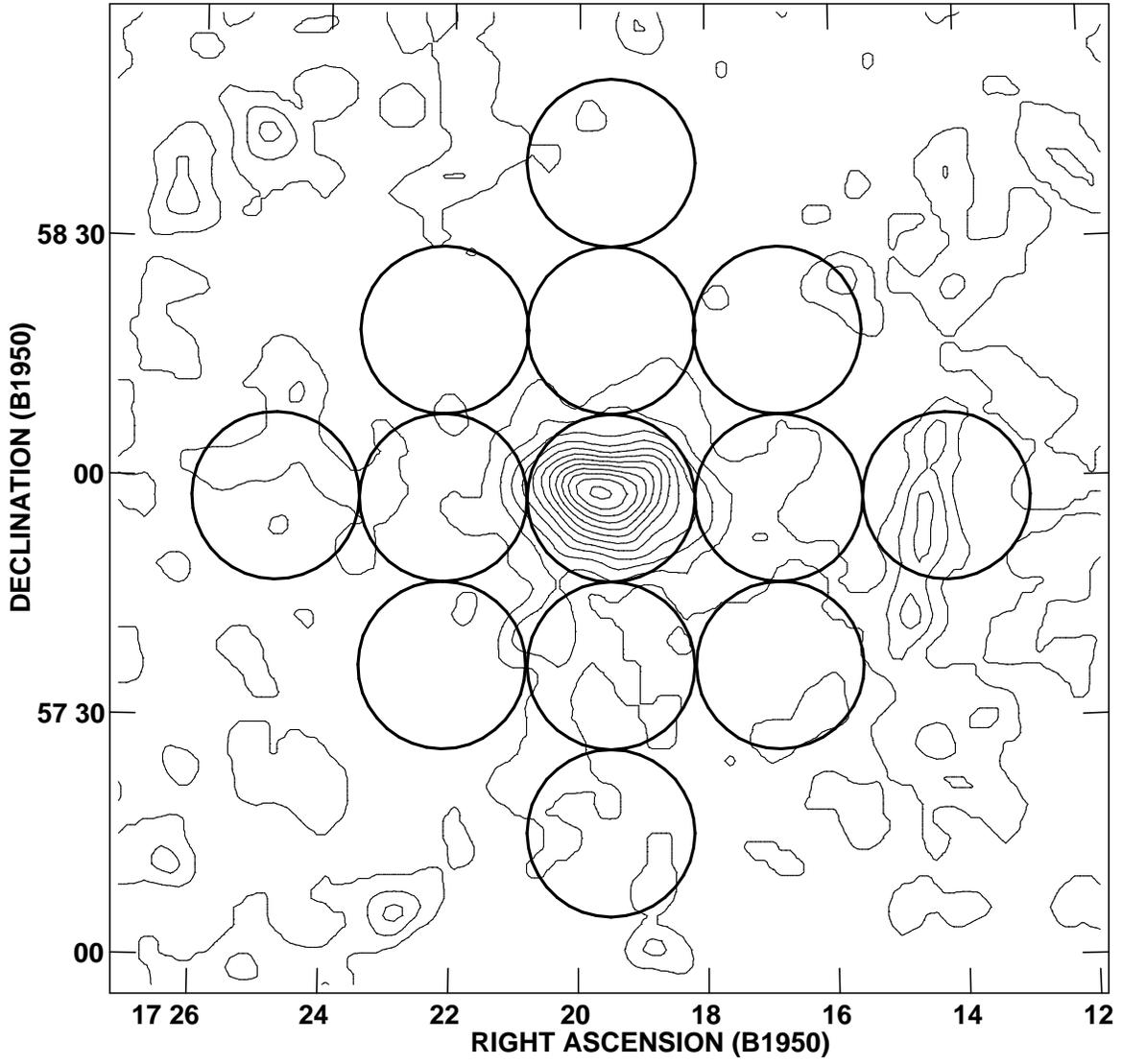

Fig. 4.— Same as Figure 1, for Draco. The most distant observations are $42'$ from the galaxy center, which is 4.7 times the core radius or 1.5 times the tidal radius.

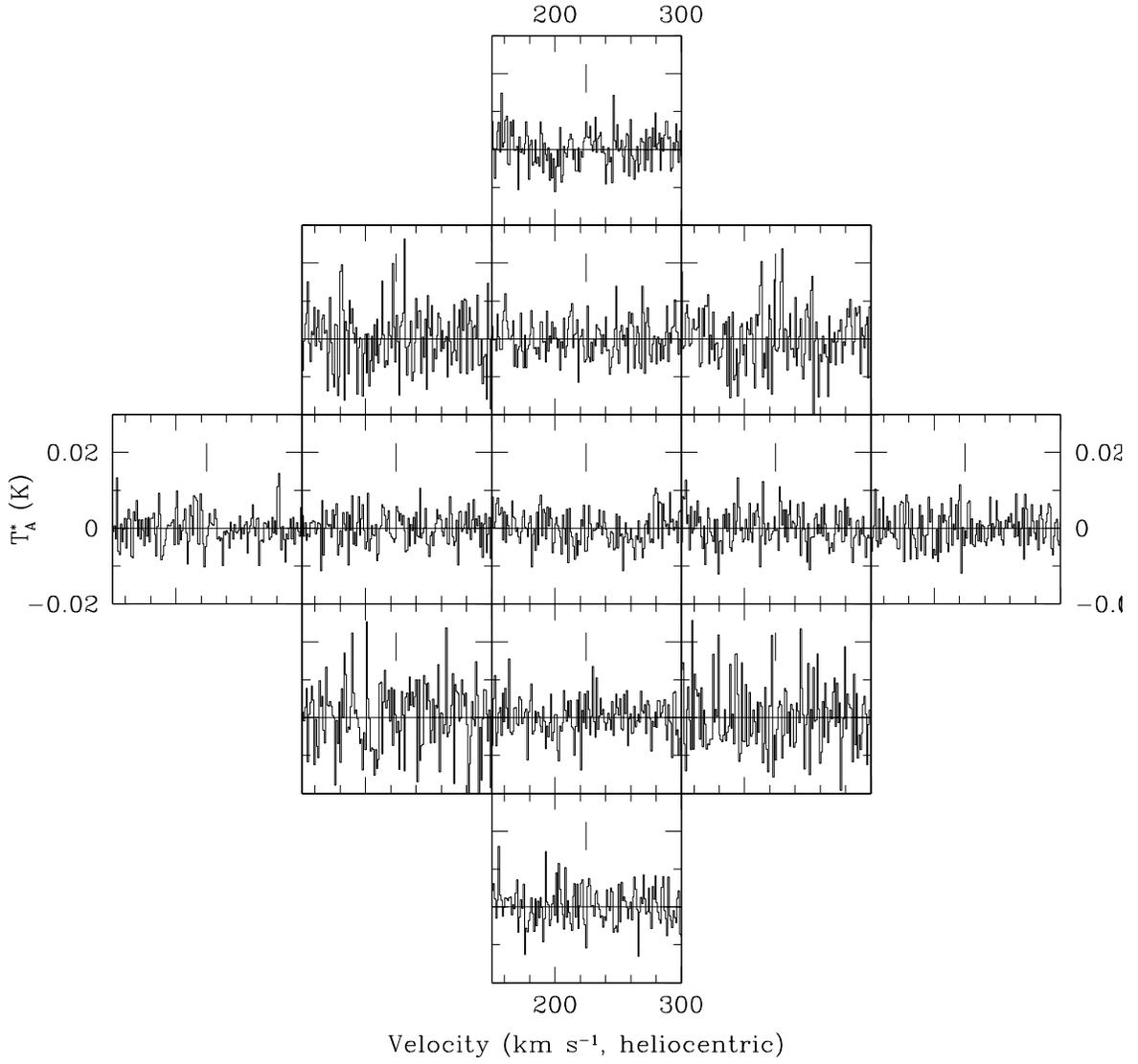

Fig. 5.— HI spectra in and around the Sextans dwarf spheroidal galaxy. The optical velocity of the galaxy is indicated with a vertical line. The figure shows only the central portion of the observed velocity range (see Table 1).

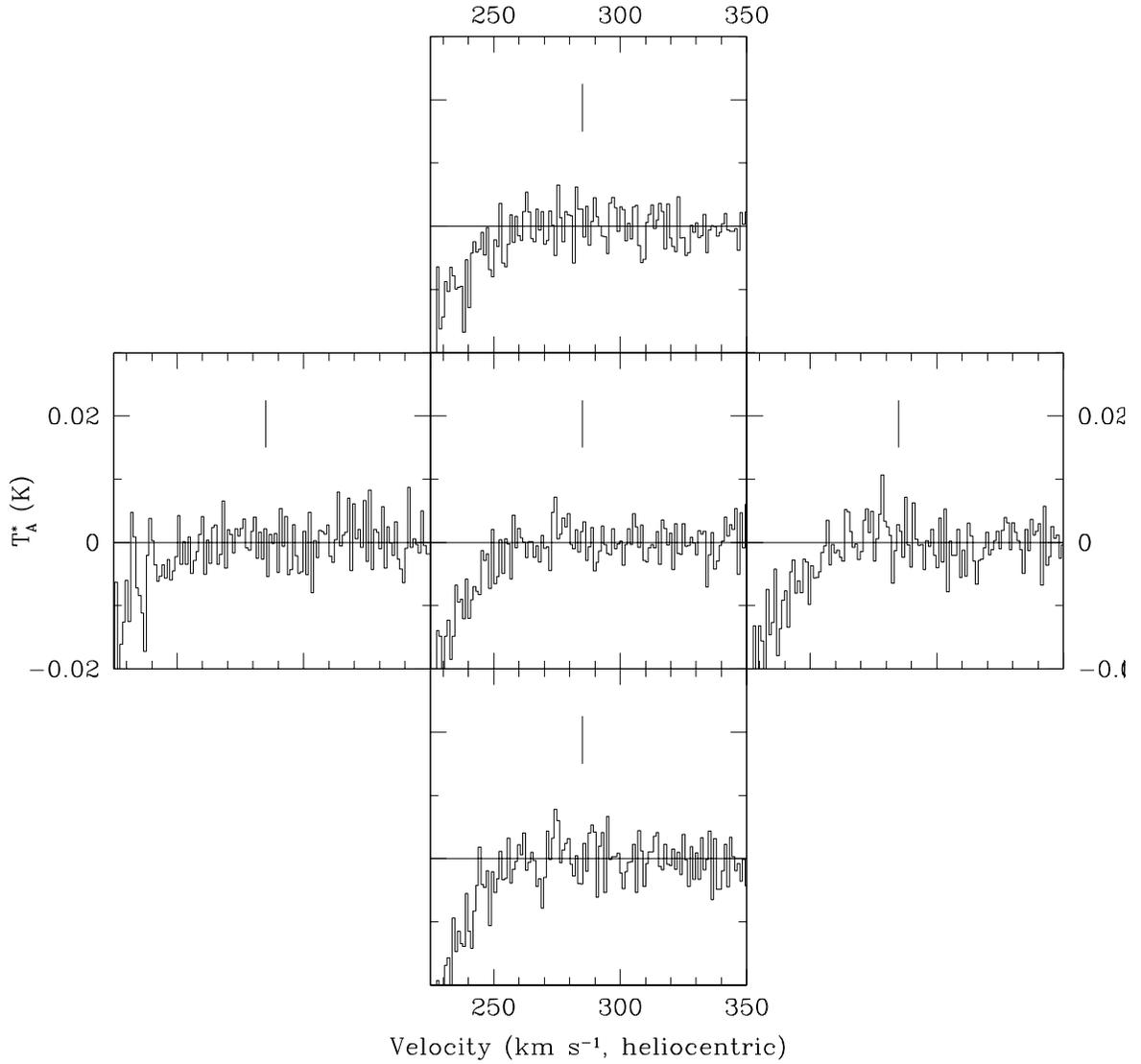

Fig. 6.— HI spectra in and around the Leo I dwarf spheroidal. As for Figure 5, the optical velocity is indicated with a vertical line. Velocities greater than $-240$ km s$^{-1}$ are corrupted by low velocity Galactic HI which was folded in by the overlapped frequency switching technique, though the center position was also observed with a wide bandwidth and without overlapped frequency switching.

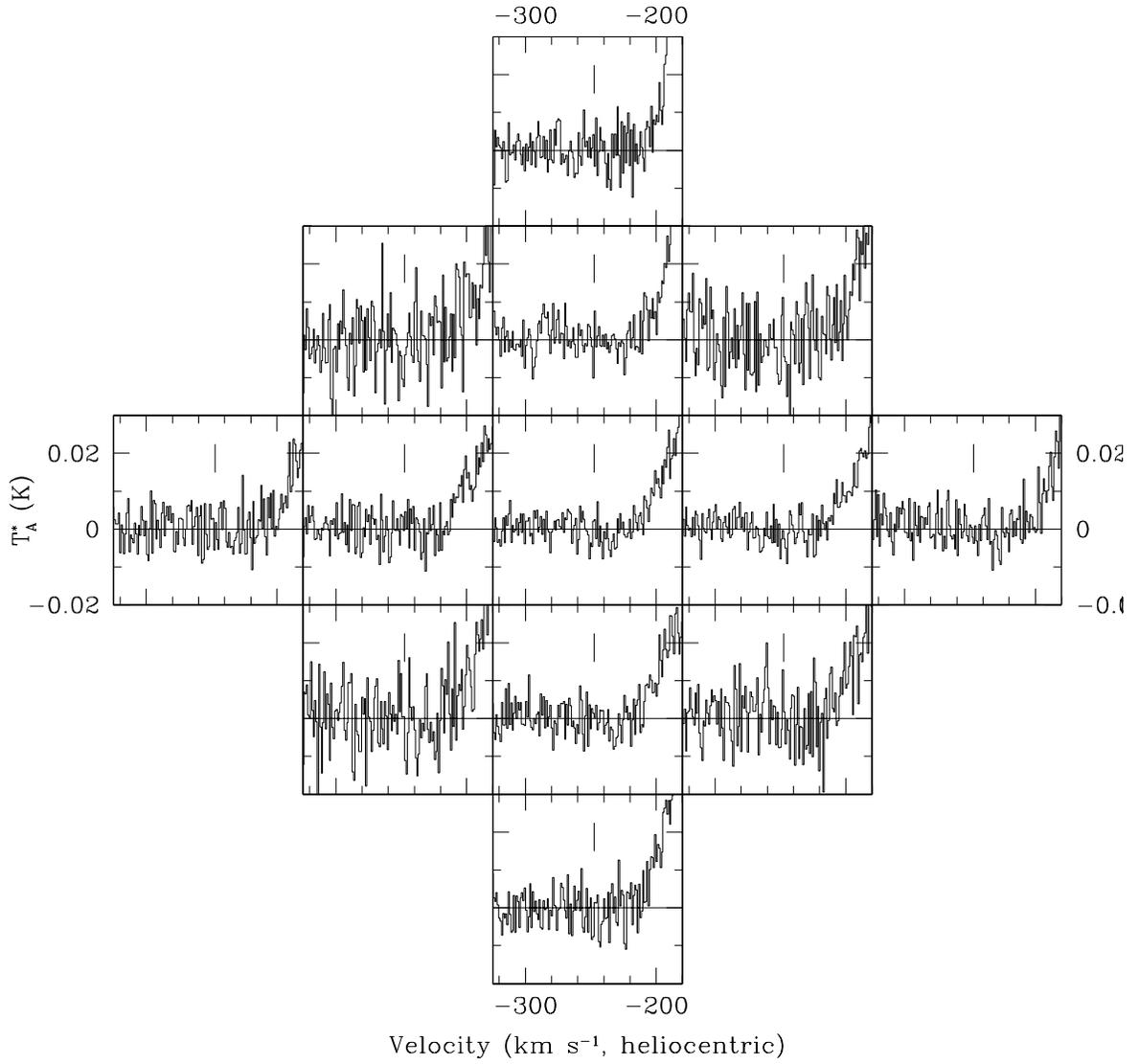

Fig. 7.— HI spectra in and around Ursa Minor. As for Figure 5.

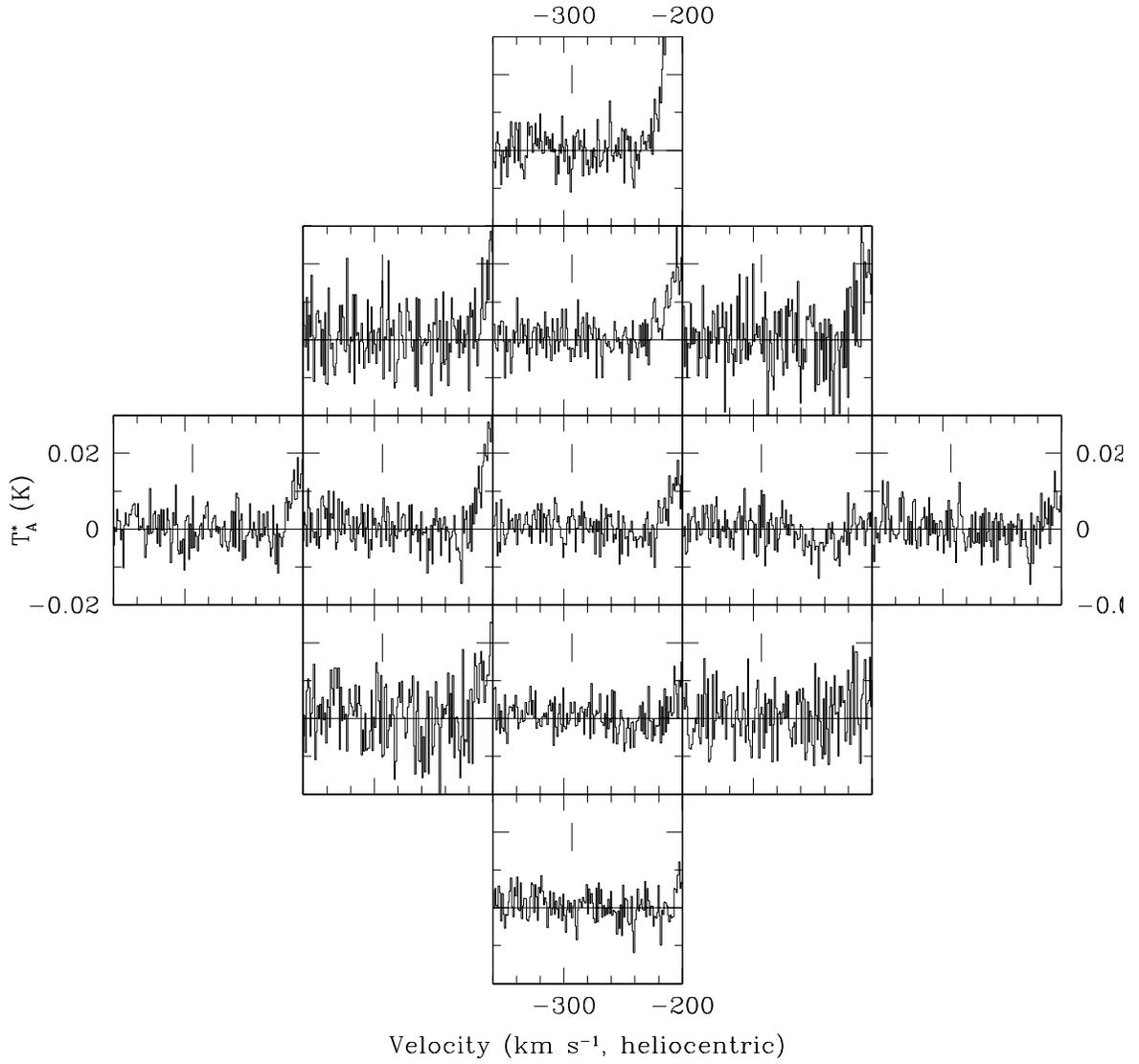

Fig. 8.— HI spectra in and around Draco. As for Figure 5.

Table 1. Properties of the observed galaxies

| Name | RA | Dec | $v_\odot$ | $r_{core}$ | $r_{tidal}$ | Distance | Gal. HI | searched |
|---|---|---|---|---|---|---|---|---|
| | B1950.0 | | km s$^{-1}$ | ′ | ′ | kpc | km s$^{-1}$ | km s$^{-1}$ |
| Ursa Minor | 15 08 34.8 | +67 24 12 | −247±1 | 15.8 | 50.6 | 64±5 | > −230 | (−470, −230) |
| Draco | 17 19 30.0 | +57 57 42 | −293±1 | 9.0 | 28.3 | 72±3 | > −220 | (−515, −220) |
| Sextans | 10 10 30.0 | −01 22 00 | +224±2 | 16.6 | 160.0 | 83±9 | <160 | (160, 460) |
| | | | | | | | | (150, 355) |
| Leo I | 10 05 46.7 | +12 33 10 | +285±2 | 3.3 | 12.6 | 198±30 | <145 | (145, 460) |
| | | | | | | | | (240, 350) |

Note. — Galaxy positions are taken from the NASA Extragalactic Database. Core and tidal radii and distances are taken from Irwin & Hatzidimitriou (1995). Heliocentric velocities $v_\odot$ are taken from various references as follows: Ursa Minor and Draco— Armandroff et al. (1995); Sextans— Hargreaves et al. (1994); Leo I— Zaritsky et al. (1989). Additional velocity references can be found in Irwin & Hatzidimitriou (1995).

Table 2.   HI Detection Limits

| Source | R.A. | Dec. | rms | Integral | N(HI) | M(HI) |
| --- | --- | --- | --- | --- | --- | --- |
| | B1950.0 | | mJy | Jy km s$^{-1}$ | $10^{17}$cm$^{-2}$ | M$_\odot$ |
| SEXTANS | 10 10 29.9 | -01 21 58.9 | 15.9 | -0.236 ± 0.137 | < 2.9 | < 670 |
| SEXTANS-N1 | 10 10 29.9 | -01 00 58.8 | 17.3 | -0.068 ± 0.157 | < 3.3 | < 770 |
| SEXTANS-S1 | 10 10 29.9 | -01 42 58.8 | 16.6 | -0.076 ± 0.150 | < 3.1 | < 730 |
| SEXTANS-E1 | 10 11 53.9 | -01 21 59.2 | 14.6 | -0.046 ± 0.133 | < 2.8 | < 650 |
| SEXTANS-W1 | 10 09 05.9 | -01 21 59.2 | 15.7 | 0.123 ± 0.142 | < 3.0 | < 690 |
| SEXTANS-N2 | 10 10 29.9 | 00 39 59.5 | 15.8 | -0.081 ± 0.143 | < 3.0 | < 700 |
| SEXTANS-S2 | 10 10 29.8 | -02 03 59.6 | 17.5 | 0.108 ± 0.159 | < 3.3 | < 780 |
| SEXTANS-E2 | 10 13 17.9 | -01 21 59.6 | 15.0 | -0.008 ± 0.136 | < 2.8 | < 660 |
| SEXTANS-W2 | 10 07 41.9 | -01 21 58.6 | 16.1 | -0.059 ± 0.146 | < 3.0 | < 710 |
| SEXTANS-NE | 10 11 53.9 | -01 00 59.6 | 27.8 | -0.106 ± 0.251 | < 5.2 | < 1230 |
| SEXTANS-SW | 10 09 05.9 | -01 42 59.5 | 28.6 | -0.058 ± 0.259 | < 5.4 | < 1270 |
| SEXTANS-SE | 10 11 53.9 | -01 42 58.8 | 30.2 | -0.050 ± 0.274 | < 5.7 | < 1340 |
| SEXTANS-NW | 10 09 05.9 | -01 00 59.1 | 27.7 | 0.138 ± 0.251 | < 5.2 | < 1230 |
| DRACO | 17 19 29.9 | 57 57 42.4 | 14.0 | 0.237 ± 0.121 | < 2.5 | < 450 |
| DRACO-N1 | 17 19 30.0 | 58 18 42.9 | 14.5 | 0.120 ± 0.126 | < 2.6 | < 460 |
| DRACO-S1 | 17 19 29.9 | 57 36 42.5 | 14.8 | 0.140 ± 0.129 | < 2.7 | < 470 |
| DRACO-E1 | 17 22 08.2 | 57 57 42.4 | 14.1 | 0.124 ± 0.122 | < 2.5 | < 450 |
| DRACO-W1 | 17 16 51.5 | 57 57 43.0 | 14.5 | 0.087 ± 0.127 | < 2.6 | < 470 |
| DRACO-N2 | 17 19 29.9 | 58 39 42.8 | 15.8 | 0.026 ± 0.137 | < 2.9 | < 500 |
| DRACO-S2 | 17 19 29.8 | 57 15 42.1 | 13.1 | 0.150 ± 0.114 | < 2.4 | < 420 |
| DRACO-E2 | 17 24 46.5 | 57 57 42.6 | 17.3 | -0.038 ± 0.150 | < 3.1 | < 550 |
| DRACO-W2 | 17 14 13.2 | 57 57 41.4 | 16.4 | 0.047 ± 0.142 | < 3.0 | < 520 |
| DRACO-NE | 17 22 08.2 | 58 18 42.4 | 27.3 | 0.172 ± 0.237 | < 4.9 | < 870 |
| DRACO-SW | 17 16 51.5 | 57 36 42.5 | 24.7 | 0.054 ± 0.215 | < 4.5 | < 790 |
| DRACO-SE | 17 22 08.2 | 57 36 42.4 | 27.0 | 0.082 ± 0.235 | < 4.9 | < 860 |
| DRACO-NW | 17 16 51.5 | 58 18 42.9 | 30.0 | 0.124 ± 0.261 | < 5.4 | < 960 |

Table 2—Continued

| Source | R.A. | Dec. | rms | Integral | N(HI) | M(HI) |
|---|---|---|---|---|---|---|
| | B1950.0 | | mJy | Jy km s$^{-1}$ | $10^{17}$cm$^{-2}$ | M$_\odot$ |
| URSAMINOR | 15 08 34.8 | 67 24 12.8 | 12.1 | -0.064 ± 0.106 | < 2.2 | < 310 |
| URSAMINOR-N1 | 15 08 34.8 | 67 45 13.0 | 14.1 | 0.015 ± 0.123 | < 2.6 | < 360 |
| URSAMINOR-S1 | 15 08 34.7 | 67 03 12.2 | 15.3 | -0.156 ± 0.133 | < 2.8 | < 390 |
| URSAMINOR-E1 | 15 12 13.3 | 67 24 11.9 | 14.9 | -0.013 ± 0.130 | < 2.7 | < 380 |
| URSAMINOR-W1 | 15 04 56.1 | 67 24 13.4 | 13.6 | -0.172 ± 0.119 | < 2.5 | < 350 |
| URSAMINOR-N2 | 15 08 34.7 | 68 06 12.2 | 16.7 | 0.041 ± 0.146 | < 3.0 | < 420 |
| URSAMINOR-S2 | 15 08 34.8 | 66 42 12.8 | 17.1 | -0.031 ± 0.149 | < 3.1 | < 430 |
| URSAMINOR-E2 | 15 15 51.9 | 67 24 12.2 | 17.8 | -0.050 ± 0.155 | < 3.2 | < 450 |
| URSAMINOR-W2 | 15 01 17.5 | 67 24 12.4 | 17.0 | -0.113 ± 0.149 | < 3.1 | < 430 |
| URSAMINOR-NE | 15 12 13.3 | 67 45 12.9 | 27.4 | 0.026 ± 0.239 | < 5.0 | < 690 |
| URSAMINOR-SW | 15 04 56.1 | 67 03 13.0 | 25.7 | 0.074 ± 0.224 | < 4.7 | < 650 |
| URSAMINOR-SE | 15 12 13.3 | 67 03 12.2 | 28.6 | -0.159 ± 0.250 | < 5.2 | < 730 |
| URSAMINOR-NW | 15 04 56.1 | 67 45 12.9 | 28.0 | -0.248 ± 0.245 | < 5.1 | < 710 |
| LEOI | 10 05 46.6 | 12 33 11.2 | 11.6 | 0.007 ± 0.103 | < 2.1 | < 2850 |
| LEOI-N1 | 10 05 46.7 | 12 54 11.6 | 11.8 | 0.039 ± 0.105 | < 2.2 | < 2920 |
| LEOI-S1 | 10 05 46.7 | 12 12 10.5 | 13.6 | 0.028 ± 0.121 | < 2.5 | < 3350 |
| LEOI-E1 | 10 07 12.7 | 12 33 10.9 | 12.7 | -0.051 ± 0.113 | < 2.3 | < 3130 |
| LEOI-W1 | 10 04 20.5 | 12 33 10.9 | 13.9 | 0.023 ± 0.124 | < 2.6 | < 3430 |

Note. — The notation "N1" after a name refers to a position which is offset from the galaxy center by one beamwidth (21′) to the north. "N2" is offset two beamwidths to the north, and so on.